\documentclass{aa}
\usepackage{graphics}

\begin{document}{\bf}


     \title{Imaging of detached shells around the carbon stars\\
     \object{R~Scl} and \object{U~Ant} through scattered stellar light
     \thanks{Based on observations using the 3.6\,m telescope of the
     European Southern Observatory, La Silla, Chile.}}

     \titlerunning{The detached shells of \object{R~Scl} and \object{U~Ant}}

     \author{D. Gonzalez Delgado\inst{1} \and H. Olofsson \inst{1}
     \and H.E. Schwarz \inst{2,3} \and K. Eriksson \inst{4} \and
     B. Gustafsson \inst{4}}

     \offprints{D. Gonzalez Delgado, delgado@astro.su.se}

     \institute{Stockholm Observatory, SE-133 36 Saltsj\"obaden,
     Sweden \and Nordic Optical Telescope, Apartado 474, E-38700 Sta.
     Cruz de La Palma, Spain \and CTIO, Casilla 603, La Serena, Chile
     \and Uppsala Astronomical Observatory, Box 515, SE-751 20
     Uppsala, Sweden }

     \date{Received date; accepted date}

     \abstract{ We present the first optical images of scattered light
     from large, detached gas/dust shells around two carbon stars,
     \object{R~Scl} and \object{U~Ant}, obtained in narrow band
     filters centred on the resonance lines of neutral K and Na, and
     in a Str{\"om}gren b filter (only \object{U~Ant}).  They confirm
     results obtained in CO radio line observations, but also reveal
     new and interesting structures.  Towards \object{R~Scl} the
     scattering appears optically thick in both the K and Na filters,
     and both images outline almost perfectly circular disks with
     essentially uniform intensity out to a sharp outer radius of
     $\approx$21$\arcsec$.  These disks are larger -- by about a
     factor of two -- than the radius of the detached shell which has
     been marginally resolved in CO radio line data.  In
     \object{U~Ant} the scattering in the K filter appears to be, at
     least partially, optically thin, and the image is consistent with
     scattering in a geometrically thin ($\approx$3$\arcsec$) shell
     (radius $\approx$43$\arcsec$) with an overall spherical symmetry.
     The size of this shell agrees very well with that of the detached
     shell seen in CO radio line emission.  The scattering in the Na
     filter appears more optically thick, and the image suggests the
     presence of at least one, possibly two, shells inside the
     43$\arcsec$ shell.  There is no evidence for such a
     multiple-shell structure in the CO data, but this can be due to
     considerably lower masses for these inner shells.  Weak
     scattering appears also in a shell which is located outside the
     43$\arcsec$ shell.  The present data do not allow us to
     conclusively identify the scattering agent, but we argue that
     most of the emission in the K and Na filter images is to due to
     resonance line scattering, and that there is also a weaker
     contribution from dust scattering in the \object{U~Ant} data.
     Awaiting new observational data, our interpretation must be
     regarded as tentative.
     } 

     \maketitle

\section{Introduction}

     Low, and intermediate-mass stars populate the asymptotic giant
     branch (AGB) after central He-burning. During this stage of
     evolution they are known to experience intense mass loss in the
     form of a cool, slow, and essentially spherically symmetric wind
     (Olofsson \cite{olofsson96b}).  The mass loss eventually becomes
     so large that it, rather than the nuclear burning, determines the
     evolutionary time scale of these stars.  This limits the AGB life
     time, and e.g.  provides an explanation to the discrepancy
     between the observed and calculated luminosity functions of stars
     on the horizontal branch and on the AGB in globular clusters
     (Renzini \cite{renzini}; Iben \cite{iben}).  The mass loss also
     makes an important contribution to the enrichment of the
     interstellar medium in terms of nuclear-processed material and
     dust particles (Busso et al.  \cite{busso}).  Considering this,
     it is unfortunate that our knowledge of the details of the mass
     loss and the mechanism behind it is so limited.

     The formation of circumstellar envelopes (CSEs) of gas and dust
     around AGB stars is a consequence of the mass loss.  The main
     constituent of the outflowing gas, H$_{2}$, is generally not
     observable, despite being excited by shocks in a few particular
     circumstances, notably young post-AGB objects and planetary
     nebulae (Beckwith et al.  \cite{beckwith}; Weintraub et al.
     \cite{weintraub}).  Standard stellar atmosphere chemistry theory
     predicts that the first molecule to form as the gas cools is CO
     (Glassgold \cite{glassgold96}).  Depending on the star being
     O-rich (i.e., [C]/[O]$<$1) or C-rich ([C]/[O]$>$1) the CO
     molecules consume all the available carbon or oxygen atoms,
     respectively.  This leads to a drastically different molecular
     chemistry and dust composition in the CSEs around these two types
     of AGB stars.  Direct measures of the gas, the principal
     component in terms of mass of a CSE, are provided by radio line
     emission of these minor molecular constituents.  Of particular
     importance for the study of the mass loss properties are
     observations of CO for both O- and C-rich stars (Olofsson et al.
     \cite{olofsson96}).

     The temporal variation of the mass loss rate is to a large extent
     unknown. This applies to all time scales from the pulsation
     period to the full time scale for the AGB-phase. For the shortest
     time scales we are limited by the spatial resolution of the
     observations, while for the longest time scales we lack suitable
     observational probes.  On the intermediate time scales
     (10$^{2}$--10$^{4}$ yr) there is now growing evidence for
     substantial variations in the mass loss rate, e.g., detached CO
     and dust shells have been detected around a number of AGB and
     post-AGB stars (Olofsson et al.  \cite {olofsson}, \cite
     {olofsson96}; Lindqvist et al.  \cite{lindqvist},
     \cite{lindqvist99}; Waters et al.  \cite{waters}; Izumiura et al.
     \cite{izumiura}, \cite{izumiura97}; Hashimoto et al.
     \cite{hashimoto}; Speck et al.  \cite{speck}), and multiple-shell
     structures have been seen in scattered light towards IRC+10216
     and a number of post-AGB objects (Harpaz et al.  \cite{harpaz};
     Kwok et al.  \cite{kwok}; Sahai et al.  \cite{sahai}; Mauron \&
     Huggins \cite{mauron99}).  Interferometric observations of the CO
     shell around the carbon star \object{TT~Cyg} provide a splendid
     result (Olofsson et al.  \cite{olofsson98}, \cite{olofsson00}).
     The shell is large (radius $\approx$35\arcsec), geometrically
     thin (average width $\approx$2.5\arcsec), and remarkably
     spherical (only $\pm$3\% variation in radius).  Similar results
     have been obtained for the carbon star U~Cam (Lindqvist et al.
     \cite{lindqvist99}).  A significant fraction of the AGB-stars has
     excess emission at 60\,$\mu$m (and 100$\mu$m) in the IRAS PSC,
     which has been interpreted as a sign of detached envelopes, and
     hence variable mass loss (Willems \& de Jong \cite{willems}; Chan
     \& Kwok \cite{chan}; Zijlstra et al.  \cite{zijlstra}).  However,
     in most cases this excess can probably be attributed to
     interstellar cirrus emission (Ivezic \& Elitzur \cite{ivezic}).
     Thus, there remain only a few, clear examples of highly episodic
     mass loss on the AGB. The exact origin of the detached shells is
     still uncertain.  They could be rare examples of `thermal
     pulse'-induced mass loss eruptions (Olofsson et al.
     \cite{olofsson}; Schr\"oder et al.  \cite{schroder}), but effects
     of interacting winds cannot be excluded, with the CO shells being
     the shock zones (Olofsson et al.  \cite{olofsson00}; Steffen \&
     Sch{\"o}nberner \cite{steffen}).

     \object{R~Scl} and \object{U~Ant} are among the few carbon stars
     with extended, detached shells detected in CO radio lines
     (Olofsson et al.  \cite{olofsson}, \cite{olofsson96}).  For
     \object{R~Scl}, the CO brightness maps show a marginally resolved
     detached shell (radius $\approx$10\arcsec), which appears to have
     detached fairly recently ($\la$10$^{3}$ yr).  Detections of other
     molecules, such as HCN, CN and CS, point to a dense shell where
     these molecules are still relatively abundant and excited
     (Bujarrabal \& Cernicharo \cite{bujarrabal}; Olofsson et al.
     \cite{olofsson96}).  The CO emission from \object{U~Ant} shows
     both a `normal' CSE and a detached shell (radius
     $\approx$40\arcsec).  The latter appears to have an overall
     spherical symmetry, but its brightness distribution is not
     smooth, suggesting clumpiness at some level in the density
     distribution, as well as a marked decrease in the CO intensity
     towards the south-west.  Izumiura et al.  (\cite{izumiura97})
     deconvolved IRAS 60 and 100\,$\mu$m observations of
     \object{U~Ant}, and they inferred the existence of two extended
     dust shells.  The (uncertain) size of the inner shell, apparent
     as a brightness distribution broader than the PSF, is consistent
     with that of the detached CO shell.  The outer shell has an inner
     radius of about 3\arcmin \, and no CO counterpart.

     The interpretation of the CO emission is hampered by the fact
     that it depends on the excitation as well as on the chemistry
     (particularly the photodissociation by the interstellar radiation
     field) of the CO molecules.  Hence, the relationship between the
     density distribution (and consequently the mass loss history) and
     the CO brightness distribution is uncertain.  Also, the lack of
     an interferometer in the southern hemisphere prevents us from
     obtaining CO radio line observations of \object{R~Scl} and
     \object{U~Ant} with the high resolution achieved for the northern
     sources \object{TT~Cyg} (Olofsson et al.  \cite{olofsson98},
     \cite{olofsson00}) and \object{U~Cam} (Lindqvist et al.
     \cite{lindqvist}, \cite{lindqvist99}).  Infrared emission from
     dust is the only probe of CSEs which is not destroyed by the
     interstellar radiation field and therefore these observations
     trace the long-term history of mass loss.  However, the data are
     often of very poor spatial resolution.  Hence, there is an urgent
     need for other ways of observing the detached shells.  In this
     paper we present the first optical images of the envelopes around
     R~Scl and U~Ant, which we infer to be the result of stellar light
     scattered in atomic resonance lines and/or by dust particles.
     These high spatial resolution data will allow us to further study
     the morphology and structure of the shells.  The feasibility of
     optically observing CSEs was first shown spectroscopically by
     Bernat \& Lambert (\cite{bernat}).  They used a method whereby
     normalized on--star spectra are subtracted from off--star
     spectra, to reveal the faint circumstellar emission; this
     technique was used and developed further by Mauron \& Caux
     (\cite{mauron92}), Plez \& Lambert (\cite{plez}), and Gustafsson
     et al.  (\cite{gustafsson}).

\section{Observations and data reduction}

\subsection {Observations}

     The imaging observations presented in this paper were done
     primarily in narrow (5\,nm FWHM) filters centred on the resonance
     lines of NaI (589.0\,nm and 589.6\,nm; the D lines) and KI
     (769.9\,nm), using the EFOSC1 and EFOSC2 focal reducer cameras on
     the ESO 3.6m telescope in 1994 and 1999, respectively.  We used
     narrow filters to reduce the amount of direct photospheric light
     in the images, and thereby increase the contrast of any
     circumstellar line scattered light.  Even narrower filters were
     considered, but they would introduce new unwanted difficulties in
     the observations, such as the temperature dependence of the
     transmitted wavelengths, and wavelength shifts over the image
     that are greater than the filter widths (produced by the large
     difference in angle at which light passes through the filter from
     one side of the field to the other in a parallel beam
     instrument).  During the second observing run, we also observed
     \object{U~Ant} in other filters, which contain no strong
     resonance lines (Str\"omgren v, $\lambda$=410.0\,nm,
     $\Delta\lambda_\mathrm{f}$=18\,nm; Str\"omgren b,
     $\lambda$=468.0\,nm, $\Delta\lambda_\mathrm{f}$=20\,nm; and
     Gunn\,z, $\lambda$=900\,nm, $\Delta\lambda_\mathrm{f}$=20\,nm,
     approximately).  In Table~\ref{Runs} we summarize the
     observations done during the two observing runs.
     
    \begin{table}
    \centering
    \caption[]{Observations performed during the two observing runs [ND,
    no shell emission detected; D(1), shell emission detected during the
    first (test) run; D(2), shell emission detected during the second run]}
     \label{Runs}
       \[
        \begin{tabular}{p{0.25\linewidth}cccccc} \hline
                     &  KI   &   Na\,D  &  Str.b  &  Str.v  &  Gunn\,z \\
              \hline
              \object{R~Scl}  & D(1) & D(1) & & &      \\
              \object{U~Ant}  & D(2) & D(1,2) & D(2) & ND & ND \\
             \object{S~Sct}  & ND & ND & & & \\
             \object{U~Hya}  & ND & ND & & & \\
             \object{R~Hya}  & ND & ND & & & \\
             \object{X~TrA}  & ND & ND & & & \\
             \object{VX~Sgr}  & ND & ND & & & \\
              \hline
           \end{tabular}
        \]

     \end{table}

     The stellar fluxes turned out to be typically four orders of
     magnitude higher than the circumstellar scattered light fluxes.
     A coronograph is therefore essential, and we used an opaque dot
     placed on a glass plate in the focal plane of the EFOSC
     instruments.  By using this method, most of the photospheric
     light that is scattered in the Earth's atmosphere is avoided.
     However, even the use of a coronographic mask is not enough, and
     an additional method is required in order to subtract the
     remaining direct photospheric radiation.  This additional method
     is based on the observation of template stars, i.e., stars with
     about the same magnitude and spectral type as the target stars,
     but without CSEs.  These template star images are subtracted from
     the target star images during the reduction.  Due to atmospheric
     variations, the stellar profiles change during the night.
     Therefore, we observed several template stars well distributed
     throughout the full runs.  Standard stars were only observed
     during the second observing run.

     The first images in circumstellar scattered light were obtained
     during observations carried out in 1994 at a test run with the
     ESO 3.6m telescope and the EFOSC1 `focal reducer'-type camera.
     The pixel plate scale was 0$\farcs$61, and exposure times of only
     a few minutes were sufficient for the detections.  Three bright
     carbon stars with detached shells detected in CO radio lines
     (\object{R~Scl}, \object{U~Ant}, \object{S~Sct}) were chosen as
     targets during this test run.  Emission due to circumstellar
     scattering in extended envelopes was found for two of them,
     \object{R~Scl} and \object{U~Ant}.  These detections are,
     however, of relatively low S/N-ratio.  Around S~Sct we found,
     using a polarimetric technique, an extremely faint and diffuse
     shell at very low S/N-ratio.  This test run turned out to be the
     first success after a series of attempts in previous years with
     other telescopes [the 2.5m Nordic Optical Telescope (NOT) and the
     MPG/ESO 2.2m telescope].  Successful spectroscopic observations
     were performed with the ESO CAT/CES-system by Gustafsson et al.
     (\cite{gustafsson}) demonstrating the presence of circumstellar
     resonance line scattering around \object{R~Scl} and other stars.
     The observational problem of imaging the circumstellar shells is
     clearly their faintness, which makes the stellar light scattered
     in the Earth's atmosphere and the telescope very problematic.
     The ESO 3.6m telescope was crucial to the outcome of the project.
     Apart from a large collecting area, it has an equatorial
     mounting.  In alt-az telescopes, like the NOT and the ESO NTT,
     the rotation of the instrument during the observations smears out
     the spider diffraction pattern over the images.  This results in
     a substantial loss of azimuthal information, rendering it
     difficult to make a detection of weak emission from an extended
     envelope.

     Based on the successful observations during the test run, we
     continued in 1999 with the imaging of one of the stars,
     \object{U~Ant}, with the EFOSC2 camera.  The pixel plate scale in
     EFOSC2 is 0$\farcs$32.  We detected again the emission from the
     envelope around \object{U~Ant}, in this case in several filters
     and with much higher S/N-ratio.  During this second run, we also
     observed two stars with detached dust shells, \object{U~Hya}
     (Waters et al., \cite{waters}) and \object{R~Hya} [the only
     O-rich source with a detected detached dust shell (Hashimoto et
     al., \cite{hashimoto})], and one source with a possible detached
     dust shell, \object{X~TrA} (Izumiura et al.  \cite{izumiura95}).
     In addition, we observed \object{VX~Sgr}, an M-type supergiant.
     We were not able to detect any circumstellar emission in these
     sources.  Due to the complexity of the observations, there can be
     several reasons for this negative result.  Normal envelopes, in
     which the density peaks close to the star, are very difficult to
     detect in scattered stellar light. It is worth noting that the
     circumstellar emission observed towards IRC+10216 is due to
     scattered interstellar light, and the central star is highly
     obscured (Mauron \& Huggins \cite{mauron99}).  Shells located
     close to the star, more extended (and therefore even fainter)
     shells (e.g., in S~Sct), and the presence of the star in a
     crowded stellar field are other plausible reasons for the
     non-detections.  We postpone the discussion of the tentative
     S~Sct shell and the non-detections to a future paper.

\subsection {Data reduction}
\label{s:reduction}

     The image reduction was done in two steps using IRAF. First, an
     ordinary CCD reduction (bias subtraction, flatfield correction,
     and cosmic ray removal) was performed.  This resulted in images
     with some considerable distortions.  In particular, the
     coronographic optics at the ESO 3.6m telescope was fairly crude,
     e.g., the focal plane obscuring spot masks are made on a glass
     plate that introduced optical distortions.  Furthermore, the lack
     of an optimized apodising mask to remove the spider diffraction
     pattern on the images, resulted in two highly saturated stripes,
     corresponding to the segments of the spider, along the horizontal
     and vertical directions of the images, where the reduction is
     much more complicated.  Therefore, the circumstellar emission
     detected in the various images shown in this paper is not
     reliable along these two directions.  Some cleaning was performed
     in the images in order to remove field stars whose brightness
     could swamp the circumstellar emission.
     
\begin{figure}
   \centering
   \resizebox{\hsize}{!}{\includegraphics{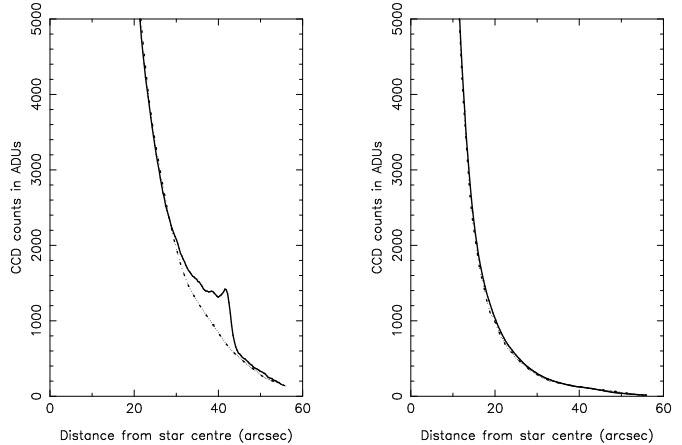}} 
   \caption{Azimuthally
   averaged radial brightness profiles of a target star (full line) and a
   template star (dot-dashed line) in the KI filter.  In the case of a
   shell detection (U~Ant, left), the target star profile has an
   additional component from the light scattered in the shell.  In the
   case of no detection (HD184835, right), the target and template star
   radial profiles match well}
\label{comprof}
\end{figure}          
    
     Second, the template star subtraction was done (the template star
     images are first reduced as above and then median filtered in
     this process).  One of the key aspects of these observations is
     the placement of the objects under the coronographic spot.  This
     is a very difficult task and when doing the reductions we have
     realized that the stars were not positioned exactly under the
     centre of the coronographic mask, and therefore the positions of
     the stars are difficult to locate precisely.  An accurate
     subtraction requires the target and template stars to be
     perfectly aligned, and so an indirect method had to be used to
     achieve this.  We estimated the positions of the stars using
     their radial profiles along various position angles.  We defined
     these positions as the middle points between two pixels (one on
     each side of the centre) with the same intensity level.  We
     averaged the values obtained using different intensity levels and
     different radial profiles (along different position angles) to
     get a final location of the target and template stars.

     Next in the reduction came the normalization of the template star
     images, i.e., normalizing the template star intensity to that of
     the target star.  Again, the absence of peaks in the images made
     this task somewhat complicated.  We normalized by a method of
     trial and error, using a radial range outside the innermost
     usable points ($\approx$7$\arcsec$ and 5$\arcsec$ from the star
     centres in the reductions of the \object{R~Scl} and
     \object{U~Ant} images, respectively) of several radial profiles.
     The profiles of the target star and the template star are usually
     perfectly matched close to the coronographic mask after
     normalization.  The addition of a small constant value to the
     template image was used to equalize the background light in the
     images due to the sky background.  The inner points, where the
     intensity is much higher, are not significantly affected by this
     addition.

     Finally, we subtracted the template star image from the target
     star image. This subtraction removes both the stellar and the
     background components from the target star image, leaving us only
     with the circumstellar emission.  We have repeated this procedure
     using two template stars, in order to make sure that the observed
     circumstellar contributions are not artifacts produced in the
     telescope or during the reduction.  As expected, in this case
     only noise is left.  Fig.~\ref{comprof} shows two examples of
     azimuthally averaged radial profiles in the KI filter: a
     circumstellar shell detection and a non-detection.  In the first
     case, the radial profile of the star (\object{U~Ant}) clearly
     shows additional emission when compared to the template star
     profile.  In the case of a non-detection, both profiles are
     essentially the same, and their subtraction produces only noise.

     Due to the fact that the centring and normalization cannot be
     performed exactly, the high values, combined with the large
     gradient, close to the star make it impossible to follow any
     shell emission inside a certain radius, and we have indicated
     this area in the final images using a central, black spot.  In
     the following we will only discuss the properties of the
     scattered emission outside this subjectively chosen area.  We
     also note here that the template star subtracted (TSS) images and
     radial profiles become progressively less reliable the closer the
     angular offset gets to this area.

     It was not possible to calibrate the \object{R~Scl} images in
     absolute fluxes because no standard stars were observed during
     the test run.  However, we have done an estimate of the shell
     fluxes by considering all possible effects which determine the
     detected number of ADUs in the CCD, see Sect.~\ref{s:fluxes}.  In
     the case of \object{U~Ant}, we have done a proper calibration of
     the shell fluxes in the resonance filters, using observations of
     a tertiary spectrophotometric standard star (LTT 3864) taken from
     the sample by Hamuy et al.  (\cite{hamuy}).

\begin{figure*}
   \centering
   \resizebox{\hsize}{!}{\includegraphics{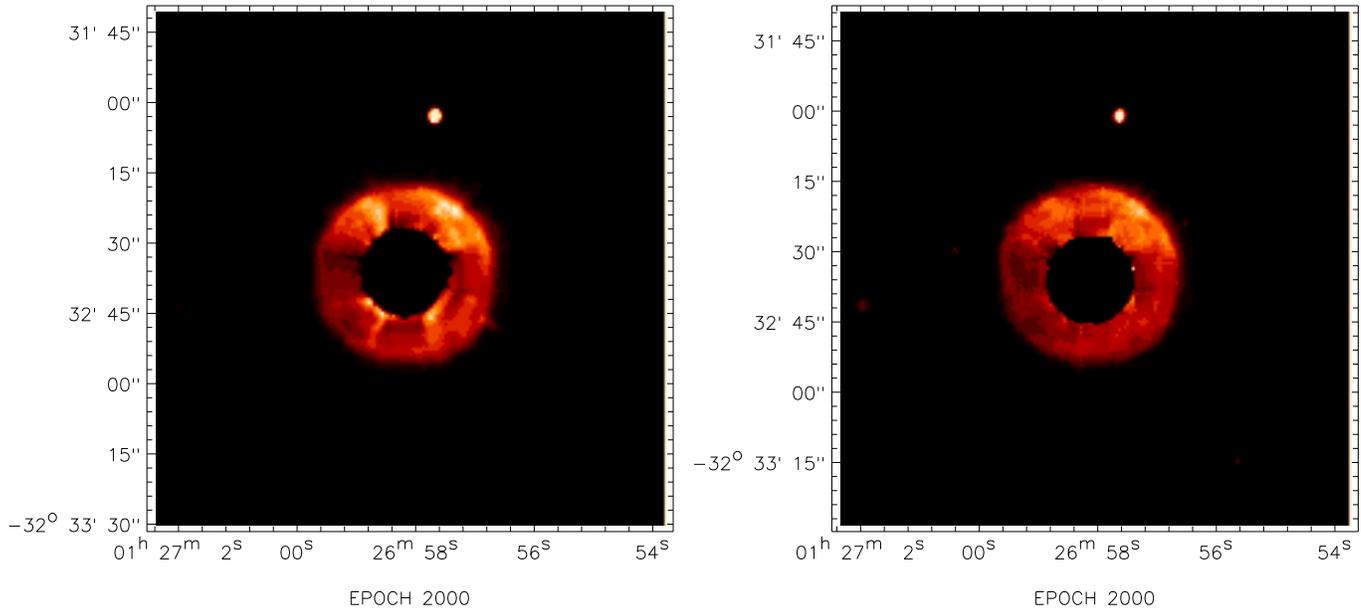}}
  \caption{Template star subtracted images of \object{R~Scl} in the
  KI (left) and Na\,D (right) filters obtained with EFOSC1.  The black
  disks mask the regions where reliable data on the scattered light does
  not exist}
   \label{rsclkna}
\end{figure*}

\section{Results}

\subsection{TSS images of \object{R~Scl} in the KI and Na\,D filters}

\begin{figure}
   \centering \resizebox{\hsize}{!}{\includegraphics{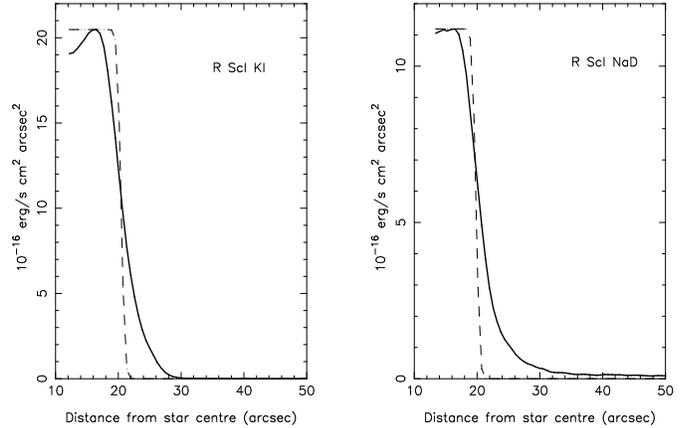}}
   \caption{Azimuthally-averaged radial brightness profiles of the
   circumstellar emission around \object{R~Scl} in the KI (left) and
   Na\,D (right) filters.  The inner parts of the profiles are
   affected by the presence of the coronographic mask and the template
   star subtraction and are therefore not shown here.  The dashed line
   shows the best-fit seeing-smoothed step function (see text for
   details)} 
\label{f:profrscl}
\end{figure}

     The TSS images of \object{R~Scl} in the KI and Na\,D filters
     reveal remaining brightness distributions in the form of
     essentially uniform-intensity disks outside the masked areas,
     Fig~\ref{rsclkna}.  The NW regions of the disks show a somewhat
     enhanced emission in both filters, but there remains distortions
     in the images that limit the useful azimuthal information.  The
     azimuthally averaged radial brightness profiles are relatively
     constant (outside the masked area) within an outer radius of
     $\approx$21$\arcsec$ in both filters (corresponding to
     1.1$\times$10$^{17}$\,cm at a distance of 360\,pc, see
     Sect.~\ref{s:structure}), Fig.~\ref{f:profrscl}.  The outer
     radius is obtained by fitting a step function, smoothed by a
     seeing Gaussian, to the radial profiles, (see
     Sect.~\ref{s:fluxes}).  The outer radius is defined as the half
     power radius of the fit.  The outer edges of the radial
     brightnesses are not as sharp as the seeing smoothed step
     function, but the half power radii are only larger by
     $\approx$2$\arcsec$ than the radii where the intensities start to
     drop.  The emission is fainter in the Na\,D filter.  We interpret
     the relative faintness in the Na\,D image as due to the star
     being fainter in this filter, and the scattering being at least
     partially optically thick (see Sect.~\ref{s:analysis}).

\begin{figure}
   \centering
   \resizebox{\hsize}{7cm}{\includegraphics{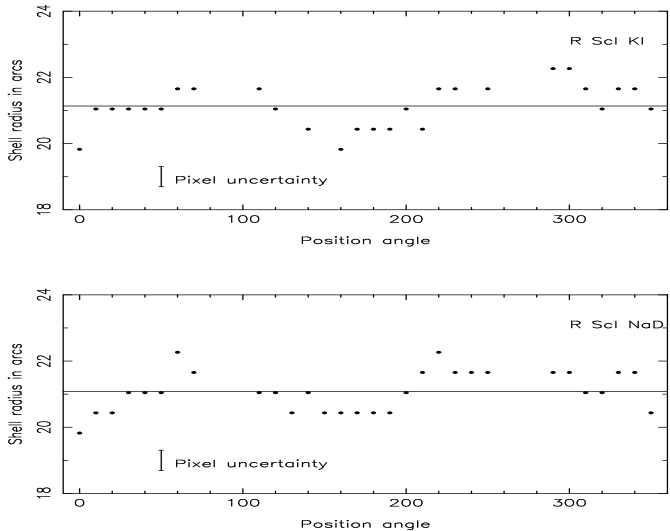}}
   \caption{Outer radii (defined in the text) of the R~Scl
   circumstellar emission in the KI (upper) and Na\,D (lower) filters
   measured at different position angles.  The thick line gives the
   mean radius obtained from these points.  The error bar indicates
   the uncertainty in the radius estimates due to the pixel size}
    \label{sphrscl}
\end{figure}

     In Fig.~\ref{sphrscl} we plot the estimated disk radii (in the KI
     and Na\,D filters), defined again as the half power radius of a
     fitted smoothed step function, as a function of position angle
     (PA).  The brightness disks appear close to circular in both
     filters, with only small deviations at the $\la$5\% level.  This
     indicates an emitting region with an overall spherical symmetry.
     The deviations can be partly attributed to the uncertainty in the
     position of the star behind the coronographic mask, but there is
     no clear tendency for an overestimate of the radii in a range of
     PAs and an underestimate in the complementary interval, as would
     be expected from a bad centring.  Furthermore, the deviations
     seem to be reproduced in the same manner in both filters,
     suggesting actual slight departures from spherical symmetry.

     Deconvolution of CO($J$=3--2) data, obtained with a 16$\arcsec$
     beam, using the Maximum Entropy Method shows a marginally
     resolved detached shell with a CO peak intensity radius which is
     significantly smaller, by about a factor of two, than the
     scattering envelope observed in the resonance line filters
     (Olofsson et al.  \cite{olofsson96}).  We will further discuss
     this in Sect.~\ref{s:structure}.

\subsection{TSS images of \object{U~Ant} in the KI and Na\,D filters}

\begin{figure*}
   \centering
   \resizebox{\hsize}{!}{\includegraphics{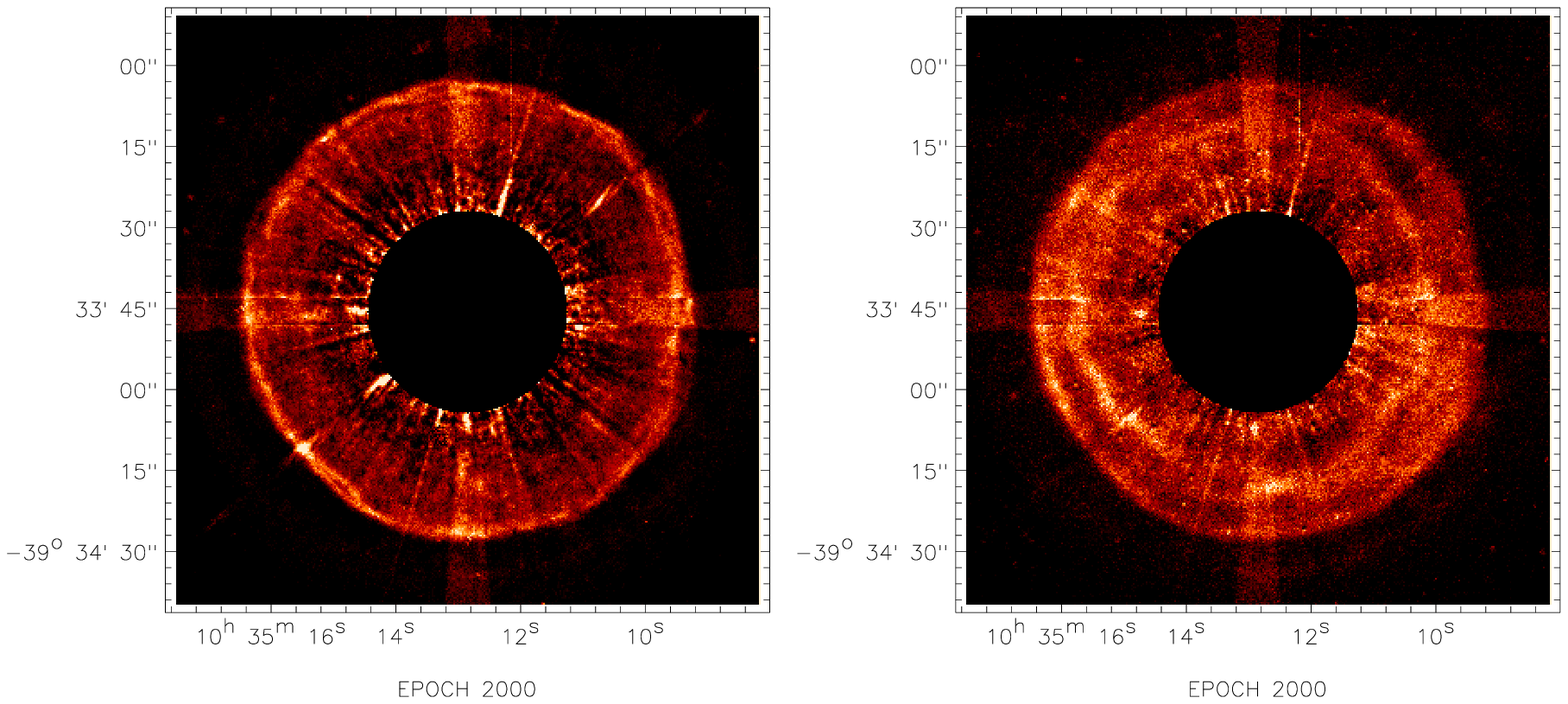}}
    \caption{Template star subtracted images of \object{U~Ant} in the
    KI (left) and Na\,D filters obtained with EFOSC2.  The black disks
    mask the regions where reliable data on the scattered light does not
    exist}
    \label{uantkna}
\end{figure*}

     The TSS images of \object{U~Ant} in the KI and Na\,D filters
     provide spectacular results, Fig~\ref{uantkna}.  In the KI filter
     image a geometrically thin, almost circular, ring with a diameter
     of $\approx$85$\arcsec$ is centred on the star.  This is most
     likely the result of limb brightening in a geometrically thin,
     essentially spherical shell.  Due to the bad coronographic
     optics, the image contains striation and diffraction spikes that
     could not be removed during the reduction, and this limits the
     available azimuthal information.
     
\begin{figure}
   \centering \resizebox{\hsize}{!}{\includegraphics{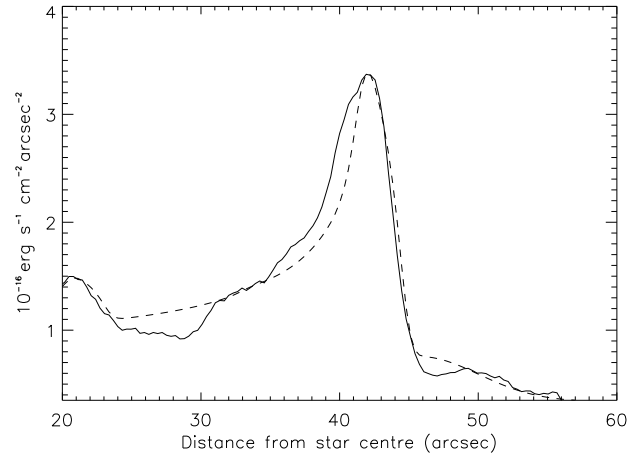}}
   \caption{Azimuthally-averaged radial brightness profile of the
   circumstellar emission around \object{U~Ant} in the KI filter.  The
   inner part of the profile is affected by the presence of the
   coronographic mask and the template star subtraction and is
   therefore not shown here.  The dashed line shows the best fit shell
   brightness distributions (see text for details)}
\label{f:profuantk}
\end{figure}

     The azimuthally-averaged radial brightness profile in the KI
     filter presented in Fig.~\ref{f:profuantk} shows a very
     pronounced intensity peak, and the mean shell radius was
     estimated to be 43$\arcsec$ (the corresponding linear radius is
     1.7$\times$10$^{17}$\,cm at the adopted distance of 260\,pc, see
     Sect.~\ref{s:structure}) by fitting a brightness distribution
     appropriate for an optically thin homogeneous shell (see
     Sect.~\ref{s:fluxes}).  This shell radius estimate from the
     scattered light agrees very well with the peak intensity radius
     of the detached CO shell observed by Olofsson et al.
     (\cite{olofsson96}), $\approx$41$\arcsec$ (they have estimated
     their angular resolution to be $\approx$15$\arcsec$, but the peak
     intensity radius has an uncertainty of only a few arc seconds).
     The emission tapers off gradually towards the centre, and there
     are indications of additional structure, which is more prominent
     in the Na\,D filter data.  The emission decreases drastically
     beyond the peak radius, but the radial profile also shows a small
     bump at $\approx$50$\arcsec$.
          
\begin{figure}
   \centering
   \resizebox{\hsize}{!}{\includegraphics{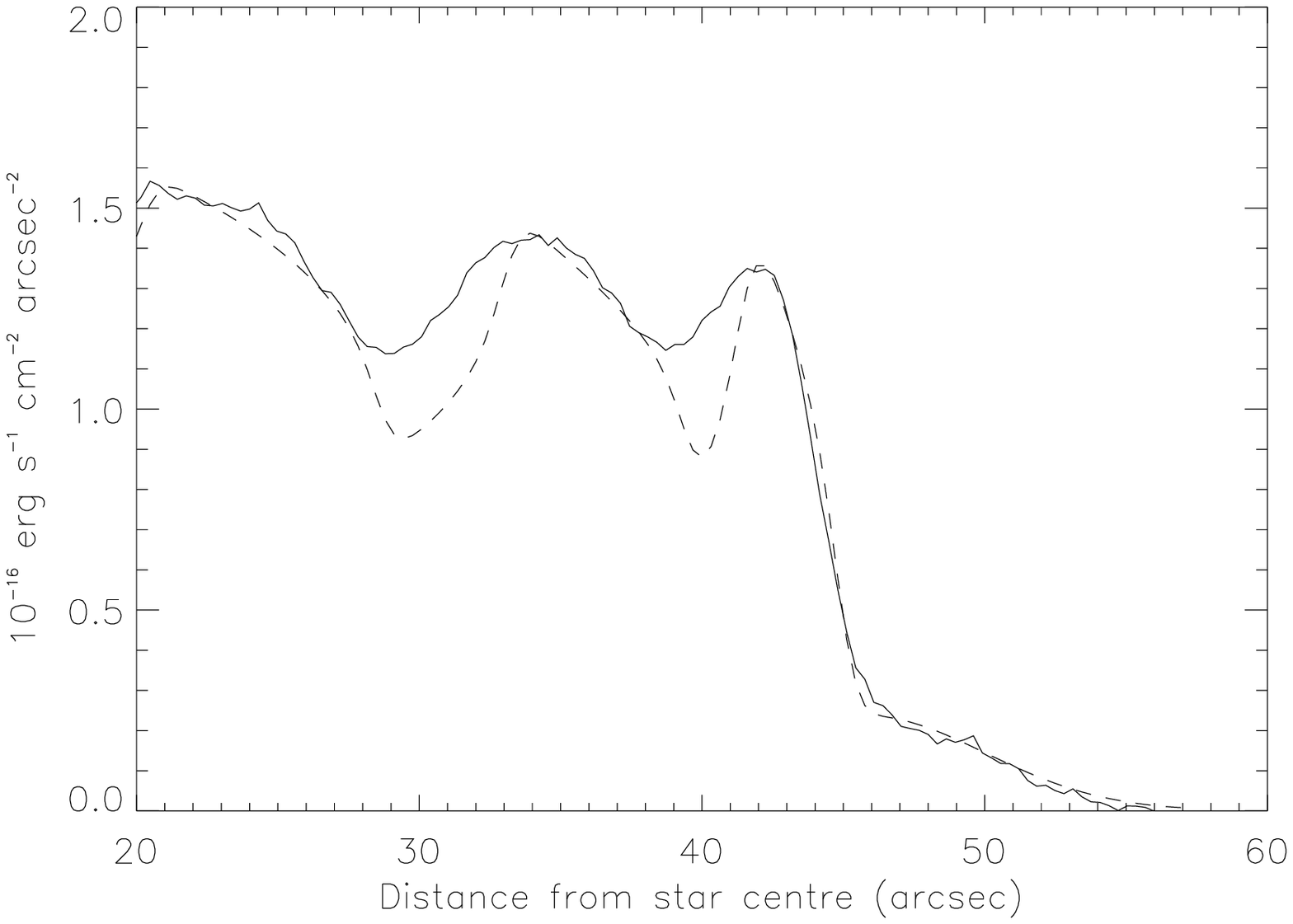}}
   \caption{Azimuthally averaged radial brightness profile of the
   circumstellar emission around \object{U~Ant} in the Na\,D filter.
   The inner part of the profiles is affected by the presence of the
   coronographic mask and the template star subtraction and is
   therefore not shown here.  The dashed line shows the best-fit shell
   brightness distributions (see text for details)}
\label{f:profuantna}
\end{figure}

     In the TSS image of \object{U~Ant} in the Na\,D filter we also
     clearly see the circumstellar emission, Fig.~\ref{uantkna}.
     However, the morphology is quite different from that obtained in
     the KI filter, in the sense that the brightness stays at nearly
     constant level much closer to the star.  In fact, the appearance
     suggests a multiple-shell structure, which is also seen in the
     azimuthally averaged radial brightness profile,
     Fig.~\ref{f:profuantna}.  At least two, maybe three, different
     components are discernible, and the two outer components may not
     be concentric.  Fitting shell brightness distributions to these
     components (see Sect.~\ref{s:fluxes}), we have determined shell
     radii of roughly 25$\arcsec$, 37$\arcsec$, and 43$\arcsec$ (1.0,
     1.4, and 1.7$\times$10$^{17}$\,cm at a distance of 260\,pc),
     although the result for the innermost component is very
     uncertain.  The outer component coincides with the dominant peak
     in the KI filter image, and the peak radius of the detached CO
     shell.  The difference between the observations in the Na\,D and
     the KI filters can be explained in terms of optical depth effects
     (see Sect.~\ref{s:analysis}).  The possible existence of multiple
     shells is surprising and this will be further discussed in
     Sect.~\ref{s:structure}. As in the KI filter image, there is a
     small bump in the radial brightness profile at
     $\approx$50$\arcsec$ offset from the star.  This emission is only
     marginally visible in the image of the shell.

\begin{figure}
   \centering
   \resizebox{\hsize}{7cm}{\includegraphics{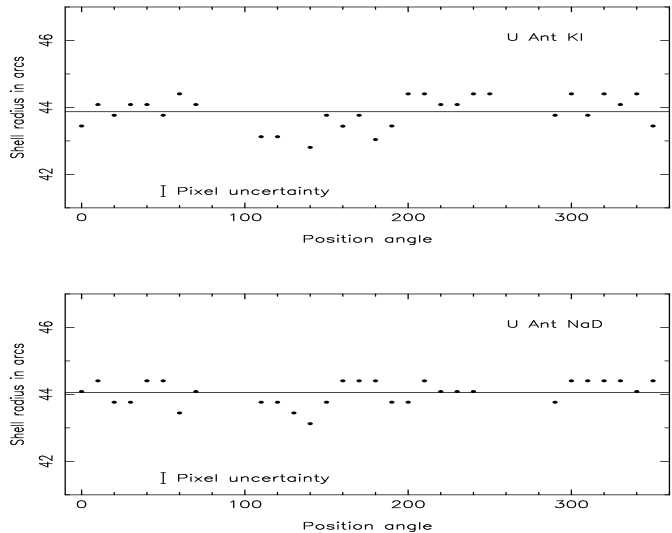}}
   \caption{U~Ant shell radii (for the 43$\arcsec$ shell) measured at
   different position angles in the KI and Na\,D filters.  The solid
   lines give the mean values for the radii obtained from these
   points.  The error bars indicate the uncertainty in the estimates
   produced by the pixel size}
\label{sphuant}
\end{figure}

\begin{figure}
   \centering \resizebox{\hsize}{!}{\includegraphics{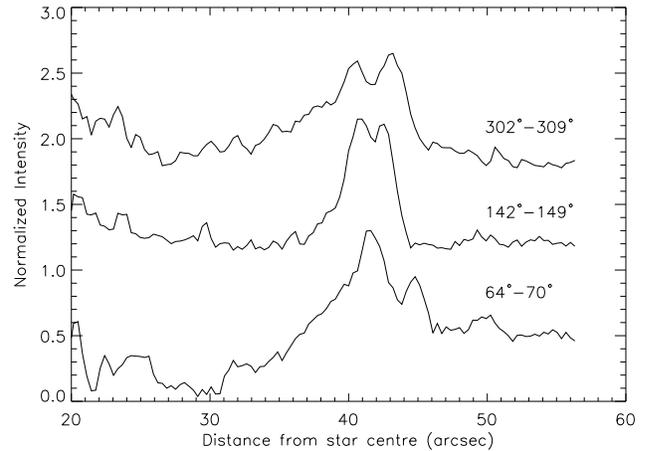}}
   \caption{Azimuthally averaged radial brightness profiles in the KI
   image of \object{U~Ant} in PA intervals where double-peaks occur.
   The profiles have been normalized to their peak values and shifted
   vertically by 0.3, 1.2, and 1.7, respectively}
\label{undulations}
\end{figure}

     Fig.~\ref{sphuant} shows the estimated shell radii (for the
     43$\arcsec$ shell) at different PAs in the KI and Na\,D filter
     images of \object{U~Ant}.  The deviations from circular symmetry
     are very small, $\la$3\%.  Thus, we can also infer a remarkable
     overall spherical symmetry for the \object{U~Ant} shell.  The
     small deviations from the means follow the same pattern in both
     filters, indicating minor irregularities in the shell morphology.
     The KI filter image reveals edge-effects, which appear in the
     form of double-peaks in the radial brightness profiles,
     Fig.~\ref{undulations}.  We interpret this as due to small
     undulations in the shell structure along the line-of-sight, but a
     more complex origin cannot be excluded.

\subsection{TSS images of \object{U~Ant} in other filters}
\label{s:uantdust}

\begin{figure}
   \centering
   \resizebox{11.5cm}{8.15cm}{\includegraphics{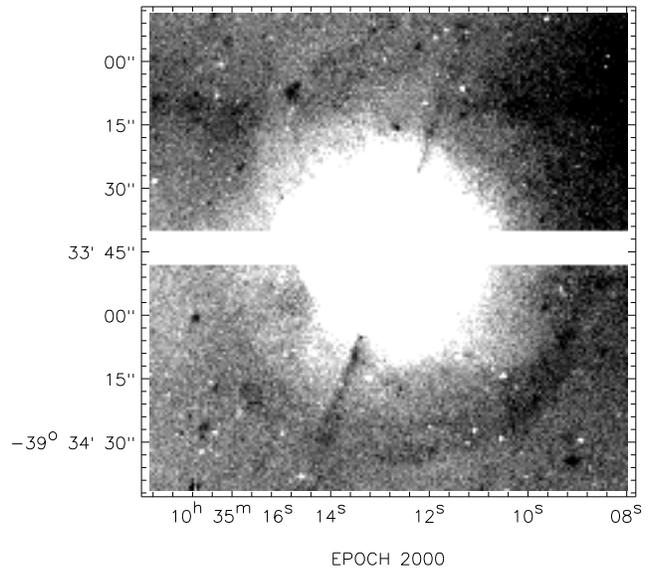}}
   \caption{Template star subtracted image of \object{U~Ant} in the b
   filter obtained with EFOSC2.  The black disk and stripe mask the
   region where reliable data on the scattered light does not exist}
   \label{uantb}
\end{figure}

     We have also made observations of \object{U~Ant} in filters
     containing no strong resonance lines (v, b, and Gunn\,z).
     Fig.~\ref{uantb} shows the TSS image of \object{U~Ant} in the b
     filter.  We have binned the image by 2$\times$2 pixels in order
     to increase the sensitivity to extended emission.  We find that
     weak emission is detected, especially towards the south.  The
     presence of a bright field star with a very long horizontal
     diffraction pattern makes the detection more problematic towards
     the north.  The azimuthally averaged radial brightness profile
     clearly shows an emission bump extending from about 35$\arcsec$
     to 55$\arcsec$.  We have estimated its radius to be
     $\approx$46$\arcsec$ by fitting a Gaussian to the radial
     brightness profile.  This is about the same radius as those of
     the small outer bumps in the KI and Na\,D radial brightness
     profiles discussed above.  We report also an upper limit,
     obtained from a short exposure, to any shell emission in the
     Gunn\,z filter.

\begin{figure}
   \centering \resizebox{\hsize}{!}{\includegraphics{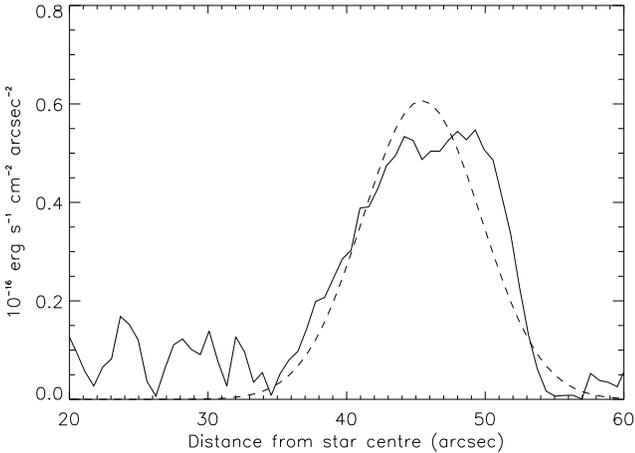}}
   \caption{Azimuthally-averaged radial brightness profile of the
   circumstellar emission around \object{U~Ant} in the b filter.  The
   inner part of the profile is affected by the presence of the
   coronographic mask and the template star subtraction and is
   therefore not shown here } 
\label{f:profuantstrb}
\end{figure}

\subsection{The stellar and circumstellar fluxes}
\label{s:fluxes}

     In order to analyse the circumstellar emission and the
     circumstellar structure it is enough to compare the relative
     fluxes of the circumstellar and stellar emission.  Unfortunately,
     the use of a coronographic mask in the observations makes it
     difficult to obtain the stellar flux from an image.  We have
     therefore tried to fit an extended point spread function of the
     type found by King (\cite{king}) and Piccirillo
     (\cite{piccirillo}) to estimate the stellar flux, but the
     resulting values appear too low by about two orders of magnitude.
     We have not identified the reason for this, but suspect that the
     use of a coronographic mask may be part of the problem.
     Therefore, an indirect method using a synthetic spectrum of
     \object{R~Scl} had to be used to estimate the stellar fluxes at
     the central wavelengths of the various filters.  We adopted
     $m_{\rm V}$=5.8 and 5.5 for \object{R~Scl} and \object{U~Ant},
     respectively. The variability of \object{R~Scl} and
     \object{U~Ant} ($\approx$0.5 mag at visual wavelengths)
     introduces some uncertainty in the calculation, but the obtained
     values should be within a factor of two of the actual fluxes.
     The results are given in Tables~\ref{rsclvalues} and
     \ref{uantvalues}.

 \begin{table*}
    \centering
    \caption[]{Results obtained from fitting a step function to the
    observed circumstellar radial brightness distributions in the
    KI and Na\,D filters of \object{R~Scl}.  The stellar fluxes are
    calculated (see text for details)}
     \label{rsclvalues}
      \[
    \begin{tabular}{p{0.08\linewidth}lcccccccc}
\hline
Filter & component & R$_{{\rm out}}$\,($\arcsec$) & $I$ (erg/s cm$^2$ $\arcsec^2$) & $F$ (erg/s cm$^2$)  & CS/S-ratio\\
\hline
KI &  shell & 21.0 & 2.0$\times$10$^{-15}$ & 2.8$\times$10$^{-12}$ &  8.5$\times$10$^{-4}$\\
   &  star  &      &                       & 3.3$\times$10$^{-9\phantom{0}}$ \\
\hline
Na\,D& shell & 20.4 & 1.1$\times$10$^{-15}$ & 1.5$\times$10$^{-12}$ &  1.1$\times$10$^{-3}$\\
     &  star &      &                      & 1.4$\times$10$^{-9\phantom{0}}$ \\
\hline    
   \end{tabular}
        \]
   \end{table*}

 \begin{table*}
    \centering
    \caption[]{Results obtained from fitting shell brightness distributions
    to the observed circumstellar radial brightness distributions in the
    KI and Na\,D filters of \object{U~Ant}.  The component external to
    the 43$\arcsec$ shell has been fitted with a Gaussian.  The stellar
    fluxes are calculated (see text for details)}
     \label{uantvalues}
   \[
	\begin{tabular}{p{0.08\linewidth}lcccccc} 
\hline
Filter & component & $R$ ($\arcsec$)  & $\Delta R$ ($\arcsec$) & $I_{\rm peak}$ (erg/s cm$^2$ $\arcsec^2$) & $F$ (erg/s cm$^2$) & CS/S-ratio\\
 \hline 
Gunn\,z & total shell &      &      & $\leq$3$\times$10$^{-16}$ & $\leq$2$\times$10$^{-12}$ & $\leq$1$\times$10$^{-4}$ \\
       & star  &      &      &                           & 2.1$\times$10$^{-8}$  &\\ 
 \hline
KI & shell1  & 25:  & 3:   & 4.3$\times$10$^{-17}$: & 5.3$\times$10$^{-14}$:& \\
 & shell2    & 36.5  & 6:  & 9.4$\times$10$^{-18}$: & 3.6$\times$10$^{-14}$:& \\
 & shell3    & 43.2 & 3.2  & 2.9$\times$10$^{-16}$  & 1.0$\times$10$^{-12}$ & 2.4$\times$10$^{-4}$\\
 & shell4    & 46.1 & 10.0 & 3.8$\times$10$^{-17}$  & 1.2$\times$10$^{-13}$ & 2.9$\times$10$^{-5}$ \\
 & total shell &      &      &                      & 1.2$\times$10$^{-12}$ & 2.9$\times$10$^{-4}$ \\
 & star      &      &      &                        & 4.2$\times$10$^{-9\phantom{0}}$  & \\
\hline
Na\,D & shell1 & 25: & 8:  & 7.6$\times$10$^{-17}$: & 1.9$\times$10$^{-13}$:& \\
 & shell2    & 36.5 & 6.4  & 9.0$\times$10$^{-17}$  & 3.5$\times$10$^{-13}$ &\\
 & shell3    & 43.4 & 3.5  & 1.2$\times$10$^{-16}$  & 4.1$\times$10$^{-13}$ & 2.3$\times$10$^{-4}$\\
 & shell4    & 46.1 & 10.0 & 2.0$\times$10$^{-17}$  & 6.2$\times$10$^{-14}$ & 3.5$\times$10$^{-5}$\\
 & total shell &      &      &                      & 1.0$\times$10$^{-12}$ & 5.6$\times$10$^{-4}$ \\
 & star      &      &      &                        & 1.8$\times$10$^{-9\phantom{0}}$  &\\ 
\hline
b   & shell4 & 46.1 & 10.0 & 5.5$\times$10$^{-18}$ & 1.8$\times$10$^{-13}$ & 3.7$\times$10$^{-4}$ \\
       & star   &      &      &                       & 4.8$\times$10$^{-10}$  &\\ 
\hline
           \end{tabular}
        \]
 \end{table*}

     The \object{U~Ant} shell fluxes (except in Gunn\,z) were
     calibrated using photometric standard stars,
     Sect.~\ref{s:reduction}.  In the case of \object{R~Scl} (and
     \object{U~Ant} in Gunn\,z) calibration data are lacking, and we
     have therefore converted from counts/s in the detector to fluxes
     above the atmosphere by correcting for atmospheric absorption,
     telescope aperture with a central obstruction, telescope
     reflectivity, instrument transmission, filter bandpasses and
     transmission, and the quantum efficiencies and gains of the CCDs.
     We estimate that the \object{R~Scl} shell fluxes are accurate to
     within a factor of five, and those of \object{U~Ant} within a
     factor of two.

     We have determined the circumstellar fluxes by fitting adopted
     brightness distributions to the azimuthally averaged radial
     brightness profiles, in order to be able to estimate the
     brightnesses in the occulted regions, and also to separate the
     fluxes from the different components.  In the case of R~Scl we
     adopt a step function, smoothed by a seeing Gaussian, to fit the
     radial profiles.  The outer radius is defined as the half power
     radius, Fig.~\ref{f:profrscl}.  In the case of U~Ant we have
     chosen, guided by the KI filter image, to assume optically thin
     scattering in homogeneous shells.  This produces a generic
     brightness distribution with the shell radius and width, and the
     peak intensity as the fitting parameters.  We have fitted such
     distributions, smoothed with a seeing Gaussian, to the three
     peaks (the components {\it shell1}, {\it shell2}, and {\it
     shell3} in Table~\ref{uantvalues}) in the Na\,D filter radial
     profile (ignoring for the moment the component outside the
     43$\arcsec$ shell).  We have done the same for the KI filter
     radial profile, except that the radii of the two inner shells
     were kept the same as for the Na\,D filter radial brightness
     profile fit.  The radial distribution of the outer component
     appears different from that of the others, and we have fitted a
     Gaussian distribution to the b filter radial brightness profile
     of U~Ant (the component {\it shell4} in Table~\ref{uantvalues}).
     A Gaussian with the same location and width was subsequently
     fitted to the outer peaks in the KI and Na\,D data.  The final
     fits are shown in Figs.~\ref{f:profuantk}, \ref{f:profuantna},
     and \ref{f:profuantstrb}.
     
     The shell surface brightnesses and fluxes obtained from the fits,
     and the ratio of circumstellar-to-stellar fluxes (CS/S-ratio) are
     given for both stars in the different filters in
     Tables~\ref{rsclvalues} and \ref{uantvalues}.  Also given are the
     estimated outer radii of the R~Scl brightness distributions, and
     the radii and widths of the U~Ant brightness distributions.

\section{Analytical approach}
\label{s:analysis}

     We will in this section derive some analytical formulae
     appropriate for a simple analysis of our observational results in
     terms of line or dust scattering, in the optically thin regime,
     of central stellar light in a geometrically thin, spherical
     shell.  Even a rough estimate shows that in our cases the stellar
     light dominates completely over the interstellar light, as
     opposed to e.g. the case of IRC+10216 which is heavily obscured
     and where the scattered light is of interstellar origin (Mauron
     \& Huggins \cite{mauron99}, \cite{mauron00}).  Clearly, the
     CS/S-ratio in a given filter cannot exceed

     \begin{equation}
     \label{e:cssthick}	 
	 \frac{F_{\mathrm{sc}}}{F_{\ast}} \leq 
	 \frac{\alpha_\mathrm{eff} \Delta \lambda_\mathrm{sc}}{\Delta\lambda_\mathrm{f}},
     \end{equation}
     where $\alpha_\mathrm{eff}$ is the ratio of the average stellar
     flux density in the total scattering wavelength range, $\Delta
     \lambda_\mathrm{sc}$, and the average stellar flux density in the
     filter passband, $\Delta\lambda_\mathrm{f}$ (e.g.,
     $\alpha_\mathrm{eff}$$<$1 in the case of resonance line
     scattering where strong photospheric line absorption removes
     photons in a wavelength range which is much narrower than the
     filter width).
     
     In the cool, detached shells of AGB stars the total line
     scattering wavelength range is determined by the turbulent motion
     of the gas, i.e., we assume there is no velocity gradient across
     the shell.  We have adopted a local velocity width of
     1\,km\,s$^{-1}$, a value which is typically assumed in the
     analysis of CSE radio line emission (Sch{\"o}ier \& Olofsson
     \cite{schoier}), and which is consistent with the CO radio line
     data of detached shells (Olofsson et al.  \cite{olofsson00}).
     This gives a maximum CS/S-ratio of about 2$\times$10$^{-4}$ and
     4$\times$10$^{-4}$ for line scattering in the narrow KI ( only
     one of the doublet lines) and Na\,D (both doublet lines) filters,
     respectively ($\alpha_\mathrm{eff}$$\approx$0.5 for the strong
     resonance lines, $\Delta\lambda_\mathrm{f}$=5\,nm).  A comparison
     with the observed values, Tables~\ref{rsclvalues} and
     \ref{uantvalues}, indicates that we are close to these ratios in
     both the KI and Na\,D filters for both sources (exceed them by at
     most a factor of four).  For dust scattering the maximum
     CS/S-ratio is 1, since $\Delta
     \lambda_\mathrm{sc}$=$\Delta\lambda_\mathrm{f}$ and
     $\alpha_\mathrm{eff}$=1.
     
      A more strict limit is provided by the fact that the observed
      surface brightness in a filter cannot exceed the source function
      for scattering at the position of the shell, i.e., the mean
      intensity of the stellar radiation and the scattered radiation
      in the optically thick shell.  This results in
     
     \begin{equation} 
	 I \leq \frac{\alpha_\mathrm{eff} F_{\ast,\lambda} 
	    \Delta \lambda_\mathrm{sc}}{\pi \theta_{\rm sh}^2},
     \end{equation}
     where $F_{\ast,\lambda}$ is the stellar flux density at the Earth
     averaged over the filter passband $\Delta\lambda_\mathrm{f}$, and
     $\theta_{\rm sh}$ is the angular size of the shell.
  
      Using the observed shell sizes the maximum surface brightnesses
      for line scattering are 5$\times$10$^{-16}$ and
      4$\times$10$^{-16}$ erg/s cm$^2$\,$\arcsec^2$ for \object{R~Scl}
      and 1$\times$10$^{-16}$ and 1$\times$10$^{-16}$ erg/s
      cm$^2$\,$\arcsec^2$ for \object{U~Ant} in the KI and Na\,D
      filters, respectively [from now on, in the case of
      \object{U~Ant}, we concentrate on the 43$\arcsec$ shell ({\it
      shell3}) for which we also have information from the CO radio
      line observations].  The observed peak surface brightnesses are
      at most a factor of four above these values.  The maximum
      surface brightnesses for dust scattering are much higher,
      2$\times$10$^{-11}$ and 1$\times$10$^{-12}$ erg/s
      cm$^2$\,$\arcsec^2$ for \object{R~Scl} in the KI and Na\,D
      filters, respectively, and 7$\times$10$^{-13}$,
      3$\times$10$^{-13}$ and 7$\times$10$^{-14}$ erg/s
      cm$^2$\,$\arcsec^2$ for \object{U~Ant} in the KI, Na\,D, and b
      filters, respectively.
     
      We will proceed with a simple analysis of the observed scattered
      emission based on the assumption of optically thin scattering in
      a shell with no velocity gradient. The circumstellar scattered
      flux at the Earth, integrated over the total scattering
      wavelength range, in the optically thin regime is given by

     \begin{equation}
         F_{\mathrm{sc},\lambda} \Delta \lambda_\mathrm{sc}
         = \frac{1}{4 \pi R_\mathrm{sh}^2} N_\mathrm{sc} \sigma_\mathrm{sc}
         \alpha_\mathrm{eff} F_{\ast,\lambda} \Delta \lambda_\mathrm{sc},
     \end{equation}
     so that the expected CS/S-ratio is given by,

     \begin{equation}
     \label{e:cssthin}	 
        \frac{F_{\mathrm{sc}}}{F_{\ast}} = \frac{F_{\mathrm{sc},\lambda} \Delta
              \lambda_\mathrm{sc}}{F_{\ast,\lambda} \Delta\lambda_\mathrm{f}} =
              \frac{\Delta \lambda_\mathrm{sc}}{\Delta\lambda_\mathrm{f}}\,
	      \frac{1}{4\pi R_\mathrm{sh}^2}
              N_\mathrm{sc} \sigma_\mathrm{sc}\alpha_\mathrm{eff},
     \end{equation}
     where $N_\mathrm{sc}$ is the number of scatterers,
     $\sigma_\mathrm{sc}$ is the scattering cross section, and
     $R_\mathrm{sh}$ is the shell radius [(\ref{e:cssthin}) is the
     optically thin limit of Eq.(\ref{e:cssthick})].  The assumption
     of optically thin scattering requires that the total scattering
     surface is smaller than the geometrical surface, i.e.,

     \begin{equation}
     \label{e:cssthinthick}	 
       \frac{N_\mathrm{sc} \sigma_\mathrm{sc}}{4\pi R_\mathrm{sh}^2} < 1\,.
     \end{equation}

\subsection{Line scattering}

     In the case of line scattering and assuming a Boltzmann
     distribution of the level populations, the number of scatterers
     and the cross section are given by,

     \begin{equation}
     \label{e:linescatterers}	 
        N_\mathrm{sc}=\frac{\eta_\mathrm{n} f_\mathrm{X} M_\mathrm{sh}}{\mu m_\mathrm{H}}
        \frac{g_{l} e^{-\frac{E_l}{kT}}}{\cal{Z}}\,,
     \end{equation}

     \begin{equation}
     \label{e:linecrossection}	 
        \sigma_\mathrm{sc}=\frac{1}{\Delta\lambda_\mathrm{sc}}\frac{1}{4
         \epsilon_{0}}
        \frac{q^2\lambda^2f}{m_\mathrm{e}c^2}\,,
     \end{equation}
     where $M_\mathrm{sh}$ is the shell mass, $\mu$ is the mean
     molecular weight, $\cal{Z}$ is the partition function, $g_l$ is
     the statistical weight of the lower level, $E_l$ is the
     excitation energy of the lower level, $f_\mathrm{X}$ is the
     number abundance of species X with respect to H,
     $\eta_\mathrm{n}$ is the fraction of X that remains in neutral
     form, and $f$ is the oscillator strength of the line (for the
     other parameters we use common notations).
     
      It is now possible to use Eqs. (\ref{e:cssthin}),
      (\ref{e:linescatterers}), and (\ref{e:linecrossection}) to
      estimate whether the observed emission in the KI and Na\,D
      filters can be attributed to line scattering.  Using the shell
      radii obtained from the observations, assuming a shell mass of
      0.01 M$_\odot$, which is a typical value derived from CO
      observations (Sch{\"o}ier \& Olofsson \cite{schoier}), and a
      solar abundance of potassium, and assuming a local velocity
      width of 1\,km\,s$^{-1}$, we find that in order to reproduce the
      observed CS/S-ratios in the KI filter only $\approx$6\% of the K
      atoms in \object{R~Scl} and $\approx$4\% in \object{U~Ant} must
      remain in neutral form.  The corresponding values assuming a
      solar abundance of sodium is $\approx$0.3\% for \object{R~Scl}
      and $\approx$0.2\% for \object{U~Ant}.
     
     The transition between optically thin and thick line scattering
     can be obtained from Eqs. (\ref{e:cssthinthick}),
     (\ref{e:linescatterers}), and (\ref{e:linecrossection}).
     Saturation in the KI line is reached with $\eta_\mathrm{n}$=0.02
     for \object{R~Scl} and $\eta_\mathrm{n}$=0.04 for \object{U~Ant}.
     For the Na\,D lines, saturation is reached already with
     $\eta_\mathrm{n}$=6$\times$10$^{-4}$ for \object{R~Scl} and
     $\eta_\mathrm{n}$=10$^{-3}$ for \object{U~Ant}.  The values of
     $\eta_\mathrm{n}$ required to obtain the observed CS/S-ratios in
     \object{R~Scl} are higher than this by a factor of a few, and
     hence any KI and Na\,D line scattering is likely to be optically
     thick.  In the case of \object{U~Ant}, at least partially,
     optically thin KI and Na\,D line scattering is possible.  We note
     here that these estimates are highly uncertain, and shall only be
     taken as a guide for the final interpretation.

     The ionization degree of the atoms in the shells is determined by
     stellar, as well as interstellar, UV photons.  If we consider the
     stars to be blackbodies with a temperature of 2500\,K and a
     luminosity of 5000\,L$_\odot$, and examine, as a first
     approximation, photons of 6\,eV (this is in the region where the
     KI photoionization cross section has its maximum, and also close
     to the ionization energy of Na), we find stellar photon flux
     densities at the shell radii of about 2$\times$10$^{-5}$
     cm$^{-2}$ s$^{-1}$ Hz$^{-1}$ in the case of \object{R~Scl} and
     10$^{-5}$ cm$^{-2}$ s$^{-1}$ Hz$^{-1}$ for \object{U~Ant} (note
     that this is very dependent on the assumed photon energy, and an
     accurate calculation requires an integration over the ionization
     cross section and the true stellar radiation field, which
     declines rapidly at these short wavelengths).  Draine
     (\cite{draine}) gives an interstellar photon flux density at 6 eV
     of about 2$\times$10$^{-3}$ cm$^{-2}$ s$^{-1}$ Hz$^{-1}$.
     Therefore, ionization by interstellar UV photons clearly
     dominates at the present time.  The atoms, however, have been
     exposed to a much stronger stellar radiation field in the past
     when they were closer to the star.  The stellar and interstellar
     photon fluxes were comparable at a radius of
     $\approx$10$^{16}$\,cm.  In other words, in R~Scl and U~Ant the
     stellar radiation dominated for about 10\% and 5\% of the
     expansion time, respectively.  The ionization is counteracted by
     recombination, but it would lead too far in this connection to
     analyse this in detail (see e.g. Glassgold \& Huggins
     \cite{glassgold}).  Mauron \& Caux (\cite{mauron92}) have
     estimated that the ionization rate of KI due to the interstellar
     radiation field is about 5$\times$10$^{-11}$ s$^{-1}$.  This
     gives an expected lifetime for the neutral species of
     $\approx$650\,yr if the interstellar radiation field dominates
     and there is no shielding.  Taken into account that both the
     \object{R~Scl} and \object{U~Ant} shells are around
     (2--3)$\times$10$^3$\,yr old, the amount of K that remains in
     neutral form in these shells is $\approx$0.02.  This number is in
     reasonable agreement with the above required values of
     $\eta_\mathrm{n}$. We expect the ionization degree of Na to be
     comparable to that of K, and hence that the Na scattering is
     considerably optically thicker than the K scattering.

     Thus, we find that the emission in the KI and Na\,D filters can,
     in principle, be explained by scattering in the resonance lines
     of those potassium and sodium atoms that remain in the detached
     shells.  Our simple calculations also suggest that the Na\,D
     lines can produce optically thick scattering (as seems to be the
     case for both \object{R~Scl} and \object{U~Ant}), while KI line
     scattering may be at least partly optically thin towards
     \object{U~Ant}, and this would explain the different appearances
     in the \object{U~Ant} KI and Na\,D filter images.

\subsection{Dust scattering}

      In the case of dust scattering, and assuming single-sized,
      spherical grains, we have the following relations for the number
      of scatterers and the scattering cross section,

     \begin{equation}
     \label{e:dustscatterers}	 
        N_\mathrm{sc}=\frac{3 f_\mathrm{dg} M_\mathrm{sh}}{4\pi \rho_\mathrm{g} a^3}\,,
     \end{equation}

     \begin{equation}
     \label{e:dustcrossection}	 
        \sigma_\mathrm{sc} = Q_\mathrm{sc} \pi a^2\,,
     \end{equation}
     where $f_\mathrm{dg}$ is the dust-to-gas mass ratio,
     $\rho_\mathrm{g}$ is the density of a dust grain, $a$ is the
     grain radius, and $Q_{\rm sc}$ is the grain scattering
     efficiency.  If the condition for Rayleigh scattering is
     fulfilled (i.e., $a/\lambda$$\ll$0.16) we have

     \begin{equation}
     \label{e:qrayleigh}	 
	 Q_{\rm sc} = \frac{128 \pi^4 a^4} {3\lambda^4} \,
	  {\rm Re} \left[ \left (\frac{\tilde n^2-1}{\tilde
	 n^2+2}\right)^2 \right]\,,
     \end{equation} 
     where $\tilde n$ is the complex index of refraction of the dust
     particle relative to the surrounding medium.
     
     Using Eqs. (\ref{e:cssthinthick}), (\ref{e:dustscatterers}), and
     (\ref{e:dustcrossection}), typical values for the different
     parameters, ($f_\mathrm{dg}$=0.005,
     $\rho_\mathrm{g}$=2\,g\,cm$^{-3}$, $a$=0.1$\mu$m, $Q_{\rm
     sc}$$<$1), and $M_\mathrm{sh}$=0.01\,M$_{\odot}$, we find for the
     shells of \object{R~Scl} and \object{U~Ant} that the area-ratios
     are $<$0.02 and $<$0.01, respectively, i.e., any dust scattering
     must be optically thin.  Alternatively, the circumstellar medium
     is highly clumped, which may indeed be the case (Olofsson et al.
     \cite{olofsson00}).
         
      We can now estimate whether it is possible to fit the
      CS/S-ratios in the KI, Na\,D, and b (for \object{U~Ant}) filters
      assuming only dust scattering.  Using Eqs. (\ref{e:cssthin}),
      (\ref{e:dustscatterers}), and (\ref{e:dustcrossection}), and the
      fact that $\Delta
      \lambda_\mathrm{sc}$=$\Delta\lambda_\mathrm{f}$, we find that
      this is, in principle, possible provided that $Q_{\rm sc}/a$ is
      about 4$\times$10$^3$\,cm$^{-1}$ and only weakly dependent on
      the wavelength.  The latter requires $a$$\geq$0.1$\mu$m, and for
      such qrains $(Q_{\rm sc}/a)$$\gg$4$\times$10$^3$\,cm$^{-1}$.  In
      addition, however, we have the requirement of optically thick
      emission for \object{R~Scl} in the KI and Na\,D filters and
      \object{U~Ant} in the Na\,D filter, which is obviously not
      consistent with the low CS/S-ratios and dust scattering, unless
      the circumstellar medium is highly clumped.
 
      We also estimate whether it is possible to explain the
      CS/S-ratios for the component {\it shell4}, which has a very
      different radial distribution than the bulk emission in the KI
      and Na\,D filters, by dust scattering.  This is possible if we
      assume a shell mass of 0.01\,M$_{\odot}$, $f_\mathrm{dg}$=0.005,
      $\rho_\mathrm{g}$=2\,g\,cm$^{-3}$, and Rayleigh scattering,
      i.e., we substitute (\ref{e:qrayleigh}) in
      (\ref{e:dustcrossection}), with an average grain size of
      0.03$\mu$m and $\tilde n$=1.6.  This grain size is consistent
      with Rayleigh scattering also in the b filter since
      $a$$\approx$0.05$\lambda$.

\section{Discussion}

\subsection{The nature of the scattered light}
\label{s:nature}
     
     No doubt, the circumstellar emission that we detect is due to
     scattering in atomic/molecular lines and/or by dust particles.
     We proceed now with a number of argument for and against
     different scattering agents.

     First, we note that the observed CS/S-ratios and surface
     brightnesses are remarkably close to the maximum allowed values
     for line scattering in the KI and Na\,D filters, and well below
     those allowed by dust scattering.  We believe that the
     uncertainties in the calibration of our data, and in some of the
     assumptions made (e.g., no velocity gradient, and the adopted
     local velocity width, shell mass, etc.), are such that a factor
     of a few can be accomodated with margin.
     
     Second, Gustafsson et al.  (\cite{gustafsson}) have shown, using
     spectroscopic observations, that in the case of \object{R~Scl}
     there exist KI resonance line scattered emission out to at least
     12$\arcsec$ from the star, and that its brightness decreases as
     the third power of the angular distance from the star, i.e., they
     see no hint of a detached shell.  However, beyond about
     12$\arcsec$ the circumstellar emission is lost in the noise;
     their CAT/CES observations are clearly less sensitive than our
     3.6m observations due to the smaller telescope and the optical
     losses in the spectrometer.  We find a brightness of about
     2$\times$10$^{-15}$\,erg s$^{-1}$ cm$^{-2}$ arcsec$^{-2}$ for the
     circumstellar emission in the KI filter beyond
     $\approx$12$\arcsec$.  This is a factor of five lower than the
     brightness obtained by Gustafsson et al.  at an offset of
     $\approx$10$\arcsec$, and hence the data are not inconsistent,
     although there is considerable uncertainty in our measured fluxes
     and to a lesser extent also in the fluxes reported by Gustafsson
     et al..
          
      Finally, the \object{U~Ant} images appear very different in the
      three filters.  In particular, the difference between the KI and
      Na\,D filter images could be naturally explained as an effect
      of, at least partially, optically thin line scattering in the
      former, and essentially optically thick line scattering in the
      latter.  Also the appearance of the \object{R~Scl} KI and Na\,D
      filter images, i.e., the uniform surface brightnesses, can only
      be explained by optically thick scattering, and since the
      observed CS/S-ratios are much less than one, this requires a
      highly clumped medium in the case of dust scattering.
     
      On the other hand, the observed emission in the b filter has a
      very different radial distribution than those in the KI and
      Na\,D filter images, and an interpretation in terms of dust
      scattering appears more reasonable here. There are reasons for
      obtaining a positive result only in the b filter.  For small
      particles the expected wavelength behaviour for the scattering
      process is such that we expect a higher contrast between the
      scattered and direct stellar components in a blue filter than in
      a red one.  For observations in a filter with a much redder
      central wavelength than that of the b filter (e.g., Gunn\,z) the
      reduction becomes more problematic, since the star is so bright
      that the circumstellar emission is swamped by the photospheric
      light even in the presence of a coronographic mask, and the
      scattering efficiency is also lower.  On the other hand, in a
      filter centred at even shorter wavelengths (e.g., v) the cool
      carbon stars have such a dramatic drop in their brightness
      (typically, B-V$\approx$3) that the number of stellar photons
      which can be scattered decreases drastically, pushing the
      circumstellar emission down below the detection limit, even if
      the dust scattering efficiency increases further at shorter
      wavelengths.
     
      Although we lack a definite proof (e.g., no spectroscopic
      information), we favour an interpretation where the main
      emission in the KI and Na\,D filters are due to resonance line
      scattering.  We further argue that weak features in the KI,
      Na\,D, and in particular b, filters are due to Rayleigh
      scattering by small grains.  Similar observations done in a
      filter containing no resonance lines and centred on the
      wavelength range between the Na\,D and KI lines (e.g.,
      H${\alpha}$) will ultimately help us discern whether the
      emission is due to resonance line or dust scattering.  Unlike
      the b and Gunn\,z filters, the stellar flux and the dust
      scattering efficiency in such a filter lie in between their
      values at the Na\,D and KI wavelengths in which we have clearly
      detected circumstellar scattered light.  Imaging through a very
      narrow filter which is tunable over the KI and/or Na\,D lines
      would also be an attractive observational mode.  Awaiting such
      data, our interpretation must be regarded as only tentative.
     
     This interpretation leads to different locations of the gas and
     the dust, in terms of the outer edges of {\it shell3} and {\it
     shell4} the difference is about 7$\arcsec$ (i.e.,
     \textit{R$_{\mathrm{dust}}$-R$_{\mathrm{gas}}$} $\approx$
     3$\times$10$^{16}$\,cm), which may at first seem surprising.
     However, this can be interpreted in terms of a drift of the dust
     particles with respect to the gas, produced by the radiation
     pressure on the dust grains.  Using a gas shell expansion
     velocity of 18.1 km s$^{-1}$ as derived from CO observations
     (Olofsson et al.  \cite{olofsson96}), this drift velocity is
     estimated to be about
     ($R_{\mathrm{dust}}-R_{\mathrm{gas}}$)$v_{\mathrm{e}}$/$R_{\mathrm{gas}}$
     $\approx$ 3\,km\,s$^{-1}$.  This is a perfectly reasonable value
     for a mass loss rate approaching
     10$^{-5}$\,M$_{\odot}$\,yr$^{-1}$ (Kwok \cite{kwok75}), which is
     in the range of mass loss rate estimates for the detached CO
     shells (Sch{\"o}ier \& Olofsson \cite{schoier}).

\subsection{The structure of the shells}
\label{s:structure}
     
     The images of R~Scl and U~Ant obtained in circumstellar scattered
     light confirm the results on detached circumstellar shells
     obtained previously from CO radio line data, but also reveal new
     interesting features.
     
     The CO data on R~Scl show a marginally resolved detached shell.
     Modelling of the CO data, as well as HCN and CN radio line data,
     suggests a relatively thick shell, $\approx$12$\arcsec$, with a
     mean radius of $\approx$12$\arcsec$, i.e., an outer radius of
     $\approx$18$\arcsec$ (Olofsson et al.  \cite{olofsson96}).  This
     outer radius agrees reasonably with the relatively sharp outer
     `disk' edges at $\approx$21$\arcsec$ in our optical images, so
     that the radio line emission and the scattered emission may be
     consistent with each other, but higher spatial resolution CO
     observations and more detailed modelling of the CO data are
     required to resolve this question.  In particular, the inability
     to separate the emission from the present mass loss envelope and
     the detached shell make the modelling of the radio lines very
     uncertain.  The high optical depths of the circumstellar
     scattered emission will make it difficult to obtain more
     information on the shell's structure from such data, although
     polarization observations may help in this respect.
     
     Additional information on the \object{R~Scl} shell is obtained
     from the spectroscopic KI observations of Gustafsson et al.
     (\cite{gustafsson}).  They found that their data, within
     10$\arcsec$ of the star, are consistent with a constant mass loss
     rate of about 5$\times$10$^{-7}$\,M$_{\odot}$ yr$^{-1}$ if 100\%
     of the potassium atoms remain neutral.  This amounts to a mass of
     about 5$\times$10$^{-4}$\,M$_{\odot}$ inside 10$\arcsec$, i.e.,
     considerably lower than the estimated CO shell mass.  This
     suggests that the shell has detached and that there is much less
     mass inside about 10$\arcsec$, but the mass estimate of
     Gustafsson et al.  can be higher if the amount of neutral
     potassium, for which there is presently no constraint, is lower.
       
     Using the expansion velocity of the detached CO shell,
     15.9\,km\,s$^{-1}$ (Olofsson et al.  \cite{olofsson96}), and a
     distance of 360\,pc derived from the period-luminosity relation
     for carbon stars (Groenewegen \& Whitelock \cite{groenewegen}),
     we estimate that the outer radius, 21$\arcsec$, corresponds to an
     age of $\approx$2.3$\times$10$^{3}$ yr.  At this time a drastic
     change in the mass loss properties of R~Scl must have occurred
     over a time scale of a few hundred years, as evidenced by the
     sharp cut-offs in the radial profiles of the scattered light.
     This conclusion is also valid in an interacting wind model, where
     the shell represents the shock zone between a fast and a slow
     stellar wind, but the time estimates may have to be redone.

     In the case of U~Ant the CO data clearly reveal a detached shell
     whose radius of $\approx$41$\arcsec$ is relatively well
     determined, but where the width is unresolved at
     $\approx$15$\arcsec$ resolution (Olofsson et al.
     \cite{olofsson96}).  To our surprise, the optical images, in
     particular the one obtained in the Na\,D filter, suggest a
     multiple-shell structure, where the shell at $\approx$43$\arcsec$
     dominates in mass, and where there is at least one inner shell,
     at $\approx$37$\arcsec$.  The dominant CO emission apparently
     comes from the 43$\arcsec$ shell.  The fit of a shell brightness
     distribution to this component in the KI filter radial brightness
     profile is not perfect, but nevertheless suggests that the shell
     is geometrically very thin, $\approx$3$\arcsec$, which should be
     regarded as an upper limit since undulations in the shell
     structure, optical depth effects, and seeing will broaden the
     azimuthally averaged radial brightness profile.  This width is
     comparable to those of the thin CO shells detected towards the
     carbon stars U~Cam and TT~Cyg (Lindqvist et al.
     \cite{lindqvist99}; Olofsson et al.  \cite{olofsson00}).  We can
     also use the fit to the KI radial brightness distribution to
     estimate the masses of the inner shells relative to that of the
     outer shell.  Taking into account that the number of photons per
     unit area to be scattered decreases as the radius squared, we
     find that the masses of the two inner shells are only a few
     percent of the mass of the outer shell.  The dynamic range of the
     CO maps is limited, and also considering the poor spatial
     resolution, we can only conclude that any CO emission from the
     two inner shells must be at least a factor of five weaker than
     that of the 43$\arcsec$ shell.  Thus, the CO data do not exclude
     the existence of inner shells with the properties found here.
     The difference between the morphology of the KI and Na\,D filter
     images can be explained by resonance line scattering and higher
     optical depths in the Na\,D lines.  This could also explain the
     poor fit of geometrically thin shell brightness distributions to
     the Na\,D data.  Optical depth effects tend to broaden the
     profile towards the star, while leaving the outer edge
     essentially unchanged.  It is not obvious why we see shells
     inside the 43$\arcsec$ shell if the Na scattering is optically
     thick.  Two possible explanations are a clumpy medium or
     differences in kinematics of the shells which will make the outer
     shell transparent to emission from the inner shells.

     Using the CO shell expansion velocity of 18.1\,km\,s$^{-1}$
     (Olofsson et al.  \cite{olofsson96}), and a distance of 260\,pc
     to \object{U~Ant} derived from the Hipparcos parallax (Alksnis et
     al.  \cite{alksnis}), the ages of the {\it shell1}, {\it shell2},
     and {\it shell3} components are estimated to be 1.7, 2.6, and
     3.0$\times$10$^{3}$ yr.  The time interval between the shells is
     consequently 400--900\,yr.  The morphology resembles to a large
     extent the multiple-shell structures (in the form of arcs, i.e.,
     incomplete shells centered on the star) seen towards the extreme
     carbon star IRC+10216 (time interval 200--800 yr; Mauron \&
     Huggins \cite{mauron99}, \cite{mauron00}), and the post-AGB
     objects IRAS 17150--3224 (200--300 yr; Kwok et al.  \cite{kwok}),
     and CRL2688 (150--450 yr; Sahai et al.  \cite{sahai}), and
     similar structures towards the M supergiant $\mu$~Cep where
     shells are seen in KI emission (Mauron \cite{mauron}).  In these
     cases the shells represent density enhancements on a general
     $r^{-2}$ density law.  Suggestions for mechanisms for the
     formation of these shell structures range from modulated mass
     loss due to a companion (Harpaz et al.  \cite{harpaz}), shock
     waves in the CSE due to a close binary (Mastrodemos \& Morris
     \cite{mastrodemos}), mass loss variations due to surface effects
     related to a magnetic activity cycle strengthened by a binary
     companion (Soker \cite{soker}), and pulsation induced effects
     internal to the star (Icke et al.  \cite{icke}).  It is not clear
     whether the putative multiple-shell structure seen towards U~Ant
     is of the same type.  An attractive alternative explanation is
     that there exist only two shells ({\it shell2} and {\it shell3}),
     of which the outer one is the shock formed by a fast, short-term
     wind running into a slower, long-term wind, and the inner shell
     is the reverse shock set up by this interaction.  The shells are
     separated by 6$\arcsec$ and over the lifetime of the outer shell
     this corresponds to a reverse velocity of about 3\,km\,s$^{-1}$,
     which is not unreasonable.  We will await more data (e.g.,
     obtained in polarization mode) before further speculating on this
     tentative multiple-shell structure and its relevance for mass
     loss on the AGB.

\acknowledgement{We are grateful to the referee, M. Jura, for detailed
and constructive criticism, which lead to a better presentation of the
data and their interpretation.  This research is partially financed by
the Swedish Natural Science Research Council, and DGD has a NOT/IAC
graduate study stipend.}

\end{document}